\documentclass[prd,aps]{revtex4}

\usepackage{epsfig}

\newcommand{\be}{\begin{equation}}
\newcommand{\ee}{\end{equation}}
\newcommand{\bea}{\begin{eqnarray}}
\newcommand{\eea}{\end{eqnarray}}

\begin{document}

\preprint{
\begin{tabular}{l}
\hbox to\hsize{\hfill KAIST-TH 2006/10}\\
[-1mm]
\hbox to\hsize{\hfill KIAS-P06047}\\
\end{tabular}
}

\title{
Exploring the charged Higgs bosons in the left-right symmetric model
}

\author{ Dong-Won Jung }
\email{dwjung@kias.re.kr}

\affiliation{
School of Physics, Korea Institute for Advanced Study,
Seoul 130-722, Korea
}

\author{ Kang Young Lee }
\email{kylee@muon.kaist.ac.kr}

\affiliation{
Department of Physics, Korea Advanced Institute of Science and Technology,
Daejeon 305-701, Korea
\\
Department of Physics, Korea University,
Seoul 136-713, Korea \footnote{Present address}
}

\date{\today}


\begin{abstract}

We explore constraints on the charged Higgs sector
in the left-right symmetric model
from the present experimental data.
Due to the different Yukawa structure,
the allowed parameter space of the charged Higgs boson
in the LR model is different from that of the two Higgs doublet model.
We find that the constraint from $t \to b H^\pm$ decay at Tevatron
is most significant, while $B \to \tau \nu$ decay could provide
no constraint on the LR model.
Bounds from $e^- e^+ \to H^- H^+$ process at LEP are similar to
those in the two Higgs doublet model.

\end{abstract}

\maketitle

\section{Introduction}

Understanding the mechanism of the electroweak symmetry breaking (EWSB)
is one of the most important motivation of new physics beyond
the standard model (SM).
The nature of the EWSB will be experimentally studied
at the CERN Large Hadron Collider (LHC) and
the $e^- e^+$ linear collider (ILC) in the future.
In the SM, one neutral Higgs boson exists as a result of the EWSB,
of which mass is not predicted in the theoretical framework.
In many models of new physics beyond the SM, more symmetries are
involved and the Higgs sector should be extended to break
larger symmetry.
Generically, the charged Higgs bosons are predicted
by models with extended Higgs sector
although they do not exist within the SM.
Thus the observation of the charged Higgs boson is
clearly a direct evidence of the new physics.
The charged Higgs boson in the two Higgs doublet (2HD) model or
the minimal supersymmetric standard model (MSSM) has been
examined through the pair production 
at the CERN Large Electron Positron Collider (LEP) \cite{lep} and
top quark decay process $t \to b H^\pm$ at Tevatron \cite{cdf}.
The absence of the observed charged Higgs boson so far
derives constraints on $(\tan \beta, m_{H^\pm})$ parameter space
for these 2HD type models.
Recent measurement of Br($B^\pm \to \tau \nu$) by Belle
provides indirect constraints \cite{btaunu}
since $B^\pm \to \tau \nu$ channel is sensitive to
the annihilation diagram mediated by the charged Higgs boson
in the 2HD model \cite{hou}.
The phenomenology of the charged Higgs boson
has been widely studied at the LHC
\cite{atlas,cms,hashemi,belyaev,bawa,barger,jung}.

Inspired by the custodial symmetry in the Higgs potential,
additional SU(2) gauge symmetry attracts much interest
as an underlying structure of the new physics
\cite{higgsless,twinhiggs}.
The left-right (LR) symmetric model based on the gauge symmetry,
SU(2)$_L \times$ SU(2)$_R \times$ U(1)$_{B-L}$,
is one of the attractive extensions of the SM
with the additional SU(2)$_R$ symmetry \cite{LR}.
In the minimal version of the LR model
with the manifest left-right symmetry,
the parity is an exact symmetry of the lagrangian
and spontaneously broken along with the gauge symmetry breaking.
The triplet Higgs boson $\Delta_R$ is introduced to break the SU(2)$_R$
symmetry and another triplet $\Delta_L$ introduced 
as a result of left-right symmetry.
The scale of the SU(2)$_R$ symmetry breaking is required
to be much higher than the electroweak scale
since the masses of the heavy gauge bosons are constrained by experiments
\cite{czagon,cheung,chay,erler}.
The right-handed fermions transform as doublets under SU(2)$_R$
and singlets under SU(2)$_L$
and the left-handed fermions behave reversely in this model.
Thus a bidoublet Higgs field is introduced
for the Yukawa couplings and the EWSB.
The triplet Higgs fields violate the lepton number and baryon number 
and do not allow the ordinary Yukawa coupling terms.
Consequently, the weak scale phenomenology of the Higgs sector 
is principally determined by the bidoublet Higgs fields,
and the dominant field contents are similar to those of the 2HD model:
a pair of the charged Higgs boson and three neutral Higgs bosons.
However, the structure of the Yukawa couplings and potential 
of the bidoublet Higgs fields are much different from
those of the 2HD model.
It leads to the different phenomenology involving charged Higgs bosons
and different constraints on the parameter space by experiments
from those in the 2HD type models.

In this paper, we examine the constraints on the charged Higgs sector
of the LR model using the present experimental results
at LEP, Tevatron and $B$-factory.
This paper is organized as follows:
In section 2, the LR model is briefly reviewed, focusing on the Higgs sector.
The analysis of $t \to b H^\pm$ process at Tevatron,
$H^\pm$ pair production at LEP,
and $B \to \tau \bar{\nu}$ decay at Belle
in the LR model are presented in section 3.
Finally we conclude in section 4.

\section{The left-right symmetric model}

The Higgs sector of the minimal LR model consists of a bidoublet
Higgs field $\phi (2, \bar{2}, 0)$ and two triplet Higgs fields
$\Delta_L (3,1,2)$ and $\Delta_R (1,3,2)$ represented by
\be
\phi =
\left(
\begin{array}{cc}
\phi_1^0 & \phi_1^+ \\
\phi_2^- & \phi_2^0 \\
\end{array}
\right),
~~~~~~~~~~~~
\Delta_{L,R} =
\frac{1}{\sqrt{2}}
\left(
\begin{array}{cc}
\delta^+_{L,R} & \sqrt{2} \delta^{++}_{L,R}  \\
\sqrt{2} \delta^0_{L,R} & -\delta^{+}_{L,R}  \\
\end{array}
\right) ,
\ee
under SU(2)$_L \times$SU(2)$_R \times$U(1)$_{B-L}$ gauge group.
The general Higgs potential has been analyzed in \cite{gunion1,gunion2,kiers}.
Minimizing condition of the Higgs potential is presented
in Eq. (10) - (15) of Ref. \cite{kiers}.
For simplicity in this work, we assume no CP violation in the Higgs sector.
The gauge symmetries are spontaneously broken by 
the vacuum expectation values (VEV)
\be
\langle \phi \rangle = \frac{1}{\sqrt{2}}\left(
\begin{array}{cc}
  k_1&0 \\
  0&k_2 \\
\end{array}
\right),
~~~~~~~~~~~~
\langle \Delta_{L,R} \rangle = \frac{1}{\sqrt{2}}\left(
\begin{array}{cc}
  0&0 \\
  v_{L,R}&0 \\
\end{array}
\right),
\ee
which lead to the charged gauge boson masses defined by
\be
{\cal L}_M = \left( W^{+ \mu}_L, W^{+ \mu}_R \right) {\cal M}^2_{W^\pm}
\left(
\begin{array}{c}
W^-_L \\
W^-_R \\
\end{array}
\right),
\ee
where 
\be 
{\cal M}^2_{W^\pm} = \frac{g^2}{4} \left(
\begin{array}{cc}
k_+^2 + 2v_L^2 &  -2 k_1 k_2 \\
-2 k_1 k_2   &  k_+^2 +2v_R^2 \\
\end{array}
\right) ,
\ee
with $k_+^2 = |k_1|^2 + |k_2|^2$.
Since SU(2)$_{\rm R}$ symmetry should be broken by $v_R$ 
at the higher scale than the electroweak scale $\sim k_+$, 
we have $k_{1,2} \ll v_R$.  
Actually $v_L$ is irrelevant for the symmetry breaking 
and just introduced in order to manifest the left-right symmetry.  
If neutrino masses are derived by the see-saw mechanism,
$m_\nu \sim v_L + k_+^2 /v_R$,
$v_R$ should be very large $\sim 10^{11}$ GeV.  
Then the heavy gauge bosons are too heavy to be produced
at the accelerator experiments
and the SU(2)$_R$ structure is hardly probed in the laboratory.
Thus we assume $v_R$ to be only moderately large for
the heavy gauge bosons to be studied at LHC.
Since $v_L$ is less than a generic neutrino mass
from the see-saw relation,
it should be very small and close to 0.
This is achieved when the quartic couplings of
$(\phi \phi \Delta_L \Delta_R)$-type terms in the Higgs potential
are set to be zero \cite{gunion2,kiers}.
We adopt this limit here.

We introduce the parameters $\xi = k_2 / k_1$ and
$\epsilon = k_1 / v_R $.
Since the parameter $\xi$ is the ratio of two VEVs for the EWSB,
it is corresponding to $\tan \beta$ in the 2HD model.
It is clear that $\epsilon \ll 1$.
We diagonalize the mass matrix of the charged gauge bosons
by a unitary transform
\be
\left(
\begin{array}{c}
W^\pm_L \\
W^\pm_R \\
\end{array}
\right)
=
\left(
\begin{array}{cc}
\cos{\zeta} & \sin{\zeta} \\
-\sin{\zeta} & \cos{\zeta} \\
\end{array}
\right)
\left(
\begin{array}{c}
W_1^\pm \\
W_2^\pm \\
\end{array}
\right),
\ee
with the mixing angle
\be
\tan{2\zeta} = \frac{2 k_1 k_2}{v_L^2 - v_R^2} \approx -2 \epsilon^2 \xi,
\ee
to yield the masses
\bea M_{W_1}^2 &\approx&
\frac{g^2}{4} |k_+|^2
    \left( 1 - \epsilon^2 \frac{2 \xi^2}{1+\xi^2} \right),
\nonumber \\
M_{W_2}^2 &\approx& \frac{g^2}{4} \cdot 2 v_R^2
    \left( 1 + \epsilon^2 \frac{1+\xi^2}{2} \right),
\eea
in the leading order of $\epsilon$.
We identify $W_1 \equiv W_{SM}$ and let $W_2 \equiv W'$ hereafter.

The Yukawa couplings for quark sector is written by
\bea
-{\cal L} = {\bar \Psi^i_L} \left( {\cal F}_{ij} \phi 
           + {\cal G}_{ij} {\tilde \phi} \right) \Psi^j_R + H.c.,
\eea
where $ \Psi^i = ( {\hat U}, {\hat D} )^\dagger$ is
the flavour eigenstates, $ {\tilde \phi} = \tau_2 \phi^\ast \tau_2$,
and ${\cal F}$, ${\cal G}$ are $3\times3$ Yukawa coupling matrices.
We rotate $ {\hat U}$ and $ {\hat D} $ into the mass eigenstates
by unitary transforms
\bea
{\hat U}_{L,R} &=& V^U_{L,R} U_{L,R},
\nonumber \\
{\hat D}_{L,R} &=& V^D_{L,R} D_{L,R},
\eea
and define Cabibbo-Kobayashi-Maskawa (CKM) matrix
$V^{CKM}_{L,R} = {V^U_{L,R}}^\dagger V^D_{L,R}$.
Here we assume the manifest left-right symmetry,
$V^{CKM}_L= V^{CKM}_R$.
Solving the equations for the diagonal mass matrices,
\bea
\frac{1}{\sqrt{2}} {V^U}^\dagger
  \left( {\cal F} k_1 + {\cal G} k_2 \right) V^U &=& {\cal M}^U,
\nonumber \\
\frac{1}{\sqrt{2}} {V^D}^\dagger
  \left( {\cal F} k_2 + {\cal G} k_1 \right) V^D &=& {\cal M}^D,
\eea
the Yukawa coupling matrices ${\cal F}$ and ${\cal G}$
are given by
\bea
{\cal F} &=& \frac{\sqrt{2}}{k_-^2}
         \left(  k_1 V^U {\cal M}^U {V^U}^\dagger
            -k_2 V^D {\cal M}^D {V^D}^\dagger \right),
\nonumber \\
{\cal G} &=& \frac{\sqrt{2}}{k_-^2}
        \left( -k_2 V^U {\cal M}^U {V^U}^\dagger
            +k_1 V^D {\cal M}^D {V^D}^\dagger \right),
\eea
where $k_-^2 = |k_1|^2 - |k_2|^2$.
If $\xi=1$, these solutions for the Yukawa coupling matrices 
given in Eq. (11) no more hold. 
Actually Eq. (10) is overdetermined and we have to treat it 
in a separate way.
On the other hand, $\xi =1$ implies the maximal LR mixing,
which is phenomenologically disfavored.
When $M_{W'}=1$ TeV, $|\tan 2 \zeta| \sim 0.01$
while the indirect constraints on the mixing angle $\zeta$
derive the bound $|\zeta| < 10^{-3}$ \cite{donoghue,wolfenstein}.
Although small $\xi$ is preferred in order to generate the ratio $m_b/m_t$, 
$\xi >1$ region cannot be excluded in general.
We do not consider the $\xi =1$ case in this work.

Taking the limit that the quartic couplings
for $(\phi \phi \Delta_L \Delta_R)$ terms and $v_L$ go to 0,
the charged Higgs boson mass matrix is given
in the basis of $(  \phi_1^+ ,\phi_2^+ ,\delta_R^+ ,\delta_L^+ )$ by
\be
{\cal M}^2_+ = \left(%
\begin{array}{cccc}
m_+^2& m_+^2 \xi & \frac{1}{\sqrt{2}} m_+^2 \epsilon (1-\xi^2)& 0\\
m_+^2 \xi & m_+^2 \xi^2 & \frac{1}{\sqrt{2}}m_+^2 \epsilon \xi (1-\xi^2) & 0 \\
\frac{1}{\sqrt{2}}m_+^2 \epsilon (1-\xi^2) & 
      \frac{1}{\sqrt{2}}m_+^2 \epsilon \xi (1-\xi^2)& 
      \frac{1}{2} m_+^2 \epsilon^2 (1-\xi^2)^2 & 0 \\
0 & 0 & 0 & m^{(+)^2}_{\rho_3} \\
\end{array}%
\right), 
\ee
where $ m_+^2 = \alpha_3 v_R^2/2 (1-\xi^2)$
with the quartic coupling $\alpha_3$ for
${\rm Tr}(\phi^\dagger \phi \Delta_L \Delta_L^\dagger)
+{\rm Tr}(\phi^\dagger \phi \Delta_R \Delta_R^\dagger)$ term
\cite{kiers}.
This limit is warranted 
by the approximate horizontal U(1) symmetry \cite{khasanov}
and the see-saw picture for light neutrino masses.  
Higgs boson masses are not affected by taking this limit \cite{kiers}.
If $\xi >1$, $\alpha_3$ should be negative to avoid 
the dangerous negative mass square of scalar fields.
Note that $\delta_L^+$ field decouples from other three charged Higgs fields
with mass $ m^{(+)^2}_{\rho_3}$
and is irrelevant for our phenomenological discussion here
since it comes from the Higgs triplet $\Delta_L$. 
By an appropriate unitary transform $V$, we diagonalize the mass matrix
$V^\dagger {\cal M}^2_+ V = {{\cal M}^2_+}^{diag}$
to lead to the transform
$(  \phi^+_1,\phi^+_2 , \delta_R^+ ,\delta_L^+ )
\rightarrow ( G_1^+,G_2^+,H^+,\delta_L^+)$,
where $G_{1,2}^+$ are Goldstone bosons and
$H^+$ is the charged Higgs boson
of which mass is given by
\be
m^2_{H^\pm} = m_+^2 (1+\xi^2)
     \left( 1+ \frac{1}{2} \epsilon^2 \frac{(1-\xi^2)^2}{1+\xi^2} \right)
     + {\cal O}(\epsilon^4),
\ee
which couples to the quarks and leptons.

Finally we obtain the $H^\pm-$quark$-$quark couplings
\be
-{\cal L} =  \bar{D} ( g_L P_L + g_R P_R) U H^- + {\rm H.c.},
\ee
where the couplings are defined by
\bea g_L &=& \sqrt{ 2 \sqrt{2} G_F} V_{UD}^{\ast}
         \left( m_U \frac{1+\xi^2}{|1-\xi^2|}
               - m_D \frac{2 \xi}{|1-\xi^2|} \right)
         \left( 1 - \frac{1}{4} \epsilon^2 (1+\xi^2) \right)
     + {\cal O}(\epsilon^4),
\nonumber \\
g_R &=& \sqrt{ 2 \sqrt{2} G_F} V_{UD}^\ast
         \left( m_U \frac{2 \xi}{|1-\xi^2|}
               - m_D \frac{1+\xi^2}{|1-\xi^2|} \right)
         \left( 1 - \frac{1}{4} \epsilon^2 (1+\xi^2) \right)
     + {\cal O}(\epsilon^4).
\eea

We also have the lepton Yukawa couplings involving the lepton number
violating terms
\bea
-{\cal L} &=&  f_{ij}{\bar \Psi^i_L} \phi \Psi^j_R 
       + g_{ij} {\bar \Psi^i_L} {\tilde \phi} \Psi^j_R + H.c. \\ \nonumber
&& + i ( h_M )_{ij}
       \left( {\Psi^i_L}^T C \tau_2 \Delta_L \Psi^j_L
          + {\Psi^i_R}^T C \tau_2 \Delta_R \Psi^j_R \right) + H.c.~,
\eea
where $f$ and $g$ are $3 \times 3$ Yukawa coupling matrices
for the Dirac masses while $h_M$ is a $3 \times 3$ Yukawa coupling matrix
to yield the lepton number violating Majorana masses.
Focusing on the charged Higgs boson coupling here,
we ignore the masses of neutrinos.
We write the Yukawa couplings for leptons by
\bea
-{\cal L}
= g_l ( {\bar l} P_L \nu_{l}) H^- + H.c.
\eea
where
\be
g_l = \sqrt{ 2 \sqrt{2} G_F} \cdot m_l \frac{2 \xi}{|1-\xi^2|}
         \left( 1 - \frac{1}{4} \epsilon^2 (1+\xi^2) \right)
     + {\cal O}(\epsilon^4).
\ee

\section{Constraints from experiments}

\subsection{Top decays at Tevatron}

The charged Higgs boson of which mass is $m_{H^\pm} < m_t-m_b$
can be produced through the top quark decay $t \to b H^\pm$
competing with the SM decay $t \to b W^\pm$.
The production of such a light charged Higgs boson has been examined
using an integrated luminosity of 193 pb$^{-1}$ data of CDF collaboration
at Tevatron in the MSSM framework \cite{cdf}.
Considering $t \bar{t}$ production,
the expected number of events for observed channel
should be modified in the presence of the charged Higgs boson,
which depends on the top and Higgs branching ratios.
Final states of $t \bar{t}$ events consist of four channels;
all-hadronic channel, lepton+jet channel, dilepton channel,
and lepton+$\tau_h$ channel.
The expected number of events in channel $k$ is given by
\be
\mu_k = \sigma_{t \bar{t}}^{\rm prod} {\cal A}_k + n_k^{\rm back},
\ee
where $\sigma_{t \bar{t}}^{\rm prod}$ is the $t \bar{t}$ production
cross section and $n_k^{\rm back}$ the number of SM-expected background events.
The detector acceptance ${\cal A}_k$ is defined by
\be
{\cal A}_k = \sum_{i,j} B_i \cdot \bar{B}_j \cdot \epsilon_{ij,k},
\ee
where $B_i$ ($\bar{B}_j$) denotes the branching ratios of
the $t$ ($\bar{t}$) quark decaying into $i(j)$-th modes,
which are
1) $t \to W^+ b$, 2) $t \to H^+ b$, $H^+ \to c \bar{s}$,
3) $t \to H^+ b$, $H^+ \to \bar{\tau} \nu$,
4) $t \to H^+ b$, $H^+ \to t^\ast \bar{b} $,
5) $t \to H^+ b$, $H^+ \to W^+ h^0, h^0 \to b \bar{b} $.
The efficiencies times integrated luminosity in channel $k$,
$\epsilon_{ij,k}$ are obtained from the Monte Carlo simulation
of $t \bar{t}$ events depending on the parameters of the model
and presented at Ref. \cite{eusebi}.
The exclusion region on the model parameter space
is obtained by the absence of the  observed charged Higgs boson
using the number of events in the CDF data,
which implies that $\Gamma(t \to H^+ b) < \Gamma_0$,
where the CDF exclusion limit $\Gamma_0$ is a function 
of model parameters.

In the LR model, the relevant model parameters are the charged Higgs
mass $m_{H^\pm}$ and the ratio of VEVs for EWSB $\xi \equiv k_2/k_1$
which is corresponding to $\tan \beta$ in the 2HD type model.
Since the dependence of the Yukawa coupling on $\xi$ is different
from that on $\tan \beta$, 
the phenomenology of the charged Higgs boson in the LR model
is also different from that in the 2HD model.
As shown in the Eq. (15), Yukawa couplings depend upon $\xi$
in the form of $(1+\xi^2)/|1-\xi^2|$ or $\xi/|1-\xi^2|$,
while those in the 2HD model are proportional to
$\tan \beta$ or $1/\tan \beta$.
Instead of using the number of events observed in the data,
we just read out the bound 
${\rm Br}|_{0} = \Gamma_0/[\Gamma_t^{\rm SM} + \Gamma(t \to H^\pm b)] $ 
by comparing the contour plot of ${\rm Br}(t \to b H^\pm)$ 
on the $(\tan \beta, m_{H^\pm})$ plane presented in Ref. \cite{eusebi}
to the exclusion limits given as the results of Ref. \cite{cdf}
for $m_{H^\pm} = 80 - 160$ GeV.
Using this bound, we vary $\xi$ and set the condition
${\rm Br}^{\rm LR}(t \to H^\pm b) < {\rm Br}|_{0} $ 
for each $m_{H^\pm}$ to obtain the exclusion limits on $(\xi, m_{H^\pm})$.
The exclusion region on $(\xi, m_{H^\pm})$ plane is depicted in Fig. 1
where the SM total decay width $\Gamma_t^{\rm SM} = 1.42$ GeV
\cite{topwidth}.
We can find that the exclusion region is much different from
that of $(\tan \beta, m_{H^\pm})$ plane in the 2HD model
presented in Ref \cite{cdf}.
We find that there is a lower bound on $m_{H^\pm} > 145$ GeV in the LR model.
Note that $m_{H^\pm}$ is proportional to $v_R$, while $m_t$
is determined by $k_+$. 
Thus the parameter region $m_{H^\pm} < m_t-m_b$ examined here 
denotes a very small $\alpha_3$ region.

\begin{center}
\begin{figure}[htb]
\hbox to\textwidth{\hss\epsfig{file=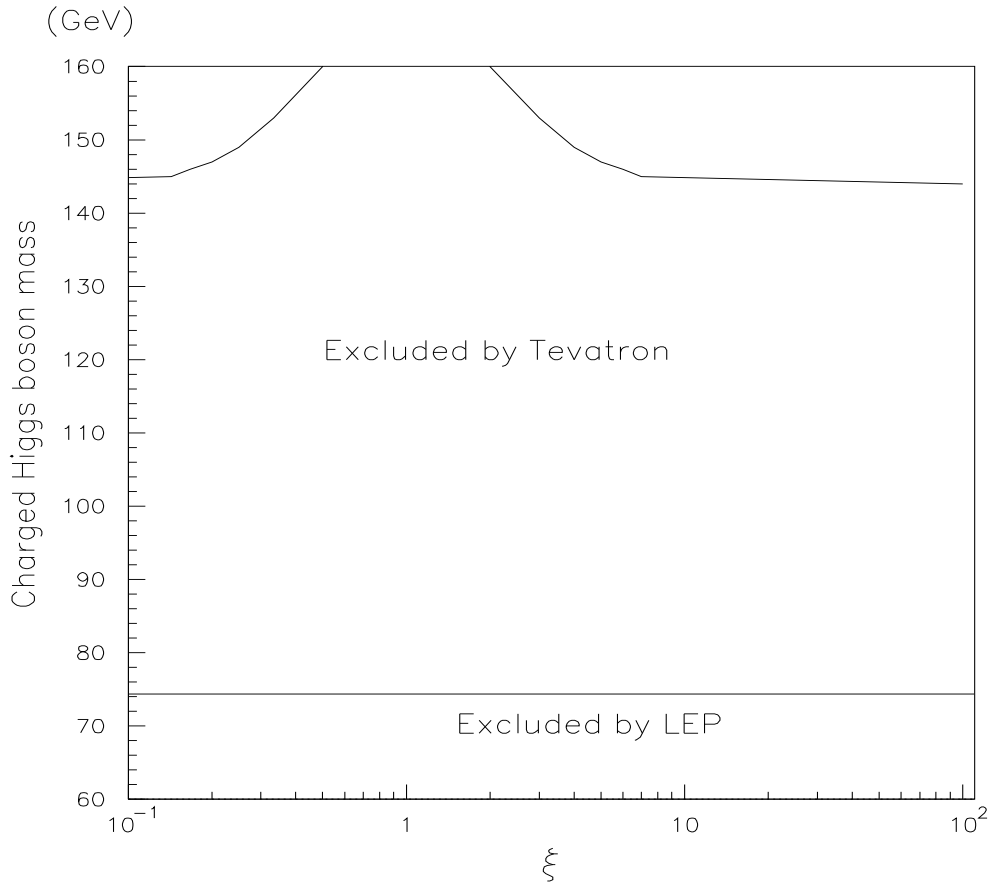,width=15cm,height=13cm}\hss}
 \vskip -1.5cm
\vspace{1cm}
\caption{
Excluded region on $(\xi, m_{H^\pm})$ parameter space
by Tevatron and LEP data at 95 $\%$ C.L..
}
\end{figure}
\end{center}

\subsection{Pair production at LEP}

At the $e^- e^+$ collider, the most promising channel of $H^\pm$
production is the tree level pair production mediated
by neutral gauge bosons.
There exist three neutral gauge fields in the LR model,
$B_\mu$, $W_L^{3 \mu}$ and $W_R^{3 \mu}$.
We diagonalize the mass matrix by an orthogonal transform
to produce one massless photon,
one massive gauge boson identical to $Z$ boson,
and one new heavy gauge boson $Z'$.
In the orthogonal transform, we need three mixing angles,
$\theta_W$, $\theta_R$, and $\theta_\xi$,
among which $\theta_W$ is identical to the Weinberg angle in the SM.
We use the notation of angles introduced in Ref. \cite{chay}
where the constraints on the mixing angles have been studied in detail.
Note that $\xi$ is replaced by $\theta_\xi$ in this paper.

We write the charged Higgs boson couplings to neutral gauge bosons 
in terms of their physical states.
The photon coupling measures the electric charge of the charged
Higgs boson and is same as that in the 2HD model.
Since the mass of $Z'$ is bounded by 630 GeV from direct search
and 860 GeV from electroweak fit \cite{pdb},
the contribution of $Z'$ is suppressed to be 10 $\%$ or less
compared with those of photon or $Z$ boson.
Ignoring the $Z'$ contribution, the cross section is given by
\cite{djouadi}
\be
\sigma (e^- e^+ \to H^- H^+) = \frac{\pi \alpha^2}{3 s}
     \left( 1 + \frac{2 g_V^e g_V^H}{1-m_Z^2/s}
        + \frac{ ({g_V^e}^2 + {g_A^e}^2) {g_V^H}^2}{(1-m_Z^2/s)^2}
     \right) \beta_H^3,
\ee
where $\beta_H = \sqrt{1-4 m_H^2/s}$ is the velocity of the Higgs boson.
The $ZH^+H^-$ coupling is expressed by the mixing of gauge bosons
and charged Higgs bosons such that
\bea
g_V^H &=& - \frac{g_L}{2} \left( \cos \theta_\xi \cos \theta_W
                   + \sin \theta_\xi \cos \theta_R
             - \cos \theta_\xi \sin \theta_W \sin \theta_R \right)
            \frac{2 (1+\xi^2)}{2(1+\xi^2) + \epsilon^2(1-\xi^2)^2}
\nonumber \\
       &&   + g_R \left( \cos \theta_\xi \sin \theta_W \cos \theta_R
                        + \sin \theta_\xi \sin \theta_R
                \right)
            \frac{\epsilon^2(1-\xi^2)^2}{2(1+\xi^2) + \epsilon^2(1-\xi^2)^2}.
\eea
Since we consider the minimal model with $g_L = g_R$, the mixing angles
satisfy the relation $ \sin \theta_R = \tan \theta_W$.
Moreover the mixing angle $\theta_\xi$ is strongly constrained
by the experiment \cite{chay}.
Thus the contribution to $g_V^H$ in the leading order of $\epsilon$
and $\theta_\xi$ is reduced to
\be
g_V^H = - \frac{e (1-2 \sin \theta_W^2)}{2 \cos \theta_W \sin \theta_W}
        + {\cal O}(\epsilon^2)
        + {\cal O}(\theta_\xi),
\ee
which is equal to that of the 2HD model.
Consequently the constraints on the charged Higgs sector
of the LR model from LEP data via the pair production
of $H^\pm$ is same in the leading order of $\epsilon$
as those in the 2HD model.
The conservative bound from LEP data is depicted in Fig. 1
by quoting in Ref. \cite{lepcn}.

\subsection{$B \to \tau \nu$ Constraints}

The purely leptonic $B \to \tau \nu$ decay proceeds via annihilation
of $B$ meson into $W$ boson in the SM
and also into $H^\pm$ in the LR model and 2HD model.
The first measurement of the branching ratio of $B \to \tau \nu$
decay has been performed by Belle \cite{btaunu}
\be
{\rm Br}(B^- \to \tau \bar{\nu}_\tau)
    = (1.79 ^{+0.56}_{-0.49} ~^{+0.39}_{-0.46}) \times 10^{-4},
\ee
which leads to the stringent constraints on the parameter set
of the 2HD model via
\be
\frac{{\rm Br}^{\rm 2HD}(B \to \tau \nu)}
     {{\rm Br}^{\rm SM}(B \to \tau \nu)}
      = \left( 1- \frac{m_B^2}{m_H^2} \tan^2 \beta \right)^2
      = 1.13 \pm 0.51,
\ee
assuming $f_B$ \cite{HPQCD} and $|V_{ub}|$ \cite{HFAG} are known.

Contributions of the LR model to $B \to \tau \nu$ decay
consist of the heavy $W$ boson and the charged Higgs boson mediation
diagrams as well as the SM contribution.
The transition amplitude is given by
\be
{\cal M} = {\cal M}_W +{\cal M}_{W'} +{\cal M}_H,
\ee
where
\bea
{\cal M}_W &=&
-\sqrt{2} G_F f_B m_\tau V_{ub} \cos \zeta (\cos \zeta + \sin \zeta)
            \left( \bar{u}(p_\tau) P_L v(p_\nu) \right),
\nonumber \\
{\cal M}_{W'} &=&
-\sqrt{2} G_F f_B m_\tau V_{ub} \sin \zeta (\sin \zeta - \cos \zeta)
     \cdot \frac{1}{2} \epsilon^2 (1+\xi^2)
            \left( \bar{u}(p_\tau) P_L v(p_\nu) \right),
\nonumber \\
{\cal M}_H &=&
- \sqrt{2} G_F f_B m_\tau V_{ub} \frac{m_B^2}{m_H^2} \frac{2 \xi}{(1+\xi)^2}
      \left( 1 - \frac{1}{2} (1+\xi^2) \epsilon^2 \right)
            \left( \bar{u}(p_\tau) P_L v(p_\nu) \right).
\eea
Since $\sin \zeta \sim \epsilon^2$, $W'$ contribution
is suppressed by $\epsilon^4$ and ignored here.
Although Yukawa couplings are proportional to $1/(1-\xi^2)$,
${\cal M}_H$ involves the combination of left-handed and right-handed
Yukawa couplings given in Eq. (15), $g_L-g_R$, which is proportional to
$(1-\xi)^2$ and cancel the divergent factor $1/(1-\xi)^2$.
The ratio of branching ratios is given by
\be
\frac{{\rm Br}^{\rm LR}(B \to \tau \nu)}
     {{\rm Br}^{\rm SM}(B \to \tau \nu)}
      = \left( 1+ \frac{m_B^2}{m_H^2} \frac{2 \xi}{(1 + \xi)^2}
                - \frac{m_W^2}{m_{W'}^2} \frac{2 \xi}{1 + \xi^2}
        \right)^2.
\ee
The new physics effects involve the factor $\xi/(1+\xi)^2$
and $\xi/(1+\xi^2)$ instead of $\tan \beta$ and $\cot \beta$.
Since these factors are at most 1 with varying $\xi$,
the new physics effects are suppressed by the mass ratio
$m_B^2/m_H^2$ and $m_W^2/m_{W'}^2$ without enhancement factors.
Consequently the LR model effects are at most a few percent
and the recent Belle measurement of Eq. (24) could not provide
any bounds on the LR model parameters.

\section{Concluding Remarks}

If we observe a charged Higgs boson, it is a clear evidence
for existence of new physics beyond the SM.
The next step is to find the underlying physics for the Higgs sector.
In this work, we examine the charged Higgs sector of the LR model
with the present experiments; $H^\pm$ pair production at LEP, top quark decay
into $b H^\pm$ at Tevatron, and $B$ annihilation decay
mediated by $H^\pm$ at $B$-factory.
Observables for each experiments depend upon
different couplings of the charged Higgs from each others.
In the 2HD model, the $t \to b H^\pm$ decay rate is governed by
the $tbH^\pm $ Yukawa coupling involving $m_t$,
which is proportional to $\tan \beta$ inversely,
while the $B \to \tau \nu$ decay depends on
the $ubH^\pm $ Yukawa coupling involving $m_b$,
proportional to $\tan \beta$.
On the other hand, the pair production of the charged Higgs bosons
at $e^- e^+$ collision is related to the gauge couplings of $H^\pm$.

In the LR model, the Yukawa couplings involve the factors of
$\xi/|1-\xi^2|$ or $(1+\xi^2)/|1-\xi^2|$ instead of
$\tan \beta$ or $1/\tan \beta$,
while the gauge couplings for the charged Higgs boson
are identical to those of the 2HD model 
in the leading order of $\epsilon = k_1/v_R$.
Thus the exclusion region of the parameter space on $(\xi, m_{H^\pm})$
in the LR model is much different from that on $(\tan \beta, m_{H^\pm})$
in the 2HD model regarding processes involving Yukawa couplings.
Constraints from the $H^\pm$ pair production through the gauge couplings
are similar in both the LR model and the 2HD model
and weaker than the Tevatron bounds obtained in this work.
The annihilation decay $B \to \tau \nu$ turns to provide
no additional constraint on the LR model parameters
because it involves the combination of left-handed and right-handed
Yukawa couplings to quarks.
In conclusion, we present the constraints on $(\xi, m_{H^\pm})$ space
for the light charged Higgs boson in the LR model
from the present experimental data.

\acknowledgments
This work was supported by
the BK21 program of Ministry of Education (K.Y.L.).

\def\PRD #1 #2 #3 {Phys. Rev. D {\bf#1},\ #2 (#3)}
\def\PRL #1 #2 #3 {Phys. Rev. Lett. {\bf#1},\ #2 (#3)}
\def\PLB #1 #2 #3 {Phys. Lett. B {\bf#1},\ #2 (#3)}
\def\NPB #1 #2 #3 {Nucl. Phys. {\bf B#1},\ #2 (#3)}
\def\ZPC #1 #2 #3 {Z. Phys. C {\bf#1},\ #2 (#3)}
\def\EPJ #1 #2 #3 {Euro. Phys. J. C {\bf#1},\ #2 (#3)}
\def\JHEP #1 #2 #3 {JHEP {\bf#1},\ #2 (#3)}
\def\IJMP #1 #2 #3 {Int. J. Mod. Phys. A {\bf#1},\ #2 (#3)}
\def\MPL #1 #2 #3 {Mod. Phys. Lett. A {\bf#1},\ #2 (#3)}
\def\PTP #1 #2 #3 {Prog. Theor. Phys. {\bf#1},\ #2 (#3)}
\def\PR #1 #2 #3 {Phys. Rep. {\bf#1},\ #2 (#3)}
\def\RMP #1 #2 #3 {Rev. Mod. Phys. {\bf#1},\ #2 (#3)}
\def\PRold #1 #2 #3 {Phys. Rev. {\bf#1},\ #2 (#3)}
\def\IBID #1 #2 #3 {{\it ibid.} {\bf#1},\ #2 (#3)}

\end{document}